\documentclass[]{meta}

\usepackage{booktabs}
\usepackage{graphicx}
\usepackage{hyperref}
\usepackage{enumitem}
\usepackage{tcolorbox}
\usepackage{pifont}
\usepackage{xcolor}
\usepackage{amsthm}

\newtheorem{definition}{Definition}
\tcbuselibrary{listings,skins,breakable}

\begin{document}
\title{Agentic Code Reasoning}

\author{Shubham Ugare}
\author{Satish Chandra}
\affiliation{Meta, USA}

\correspondence{Shubham Ugare at \email{sugare@meta.com}}



\abstract{
Can LLM agents explore codebases and reason about code semantics without executing the code? We study this capability, which we call \emph{agentic code reasoning}, and introduce \emph{semi-formal reasoning}: a structured prompting methodology that requires agents to construct explicit premises, trace execution paths, and derive formal conclusions. Unlike unstructured chain-of-thought, semi-formal reasoning acts as a certificate: the agent cannot skip cases or make unsupported claims.
We evaluate across three tasks (patch equivalence verification, fault localization, and code question answering) and show that semi-formal reasoning consistently improves accuracy on all of them.
For patch equivalence, accuracy improves from 78\% to 88\% on curated examples and reaches 93\% on real-world agent-generated patches, approaching the reliability needed for execution-free RL reward signals. For code question answering on RubberDuckBench~\cite{rubberduckbench}, semi-formal reasoning achieves 87\% accuracy, a 9 percentage point gain over standard agentic reasoning. For fault localization on Defects4J~\cite{just2014defects4j}, semi-formal reasoning improves Top-5 accuracy by 5 percentage points over standard reasoning. These results demonstrate that structured agentic reasoning enables meaningful semantic code analysis without execution, opening practical applications in RL training pipelines, code review, and static program analysis.
}

\maketitle

\section{Introduction}
Can an LLM agent explore a codebase and determine whether two patches are semantically equivalent, without ever running them? We study this capability, which we call \emph{agentic code reasoning}: an agent's ability to navigate files, trace dependencies, and gather context iteratively to perform deep semantic analysis without executing code. This capability is essential for tasks like bug detection, code review, and patch verification in complex repositories where relevant context spans multiple files.
However, evaluating whether an agent's reasoning is \emph{correct} presents a fundamental challenge: how do we establish ground truth for code understanding? And how do we ensure agents reason thoroughly rather than guess?

Recent work has explored execution-free verification for code agents. SWE-RM~\cite{shum2025swerm} trains reward models to approximate test outcomes, Agentic Rubrics~\cite{raghavendra2025rubrics} decompose verification into LLM-generated criteria, and CodeJudge~\cite{tong2024codejudge} uses LLMs directly as evaluators. However, these approaches use unstructured reasoning, allowing models to make claims about code behavior without explicit justification.
At the other extreme, formal verification approaches translate code or reasoning into formal languages like Lean~\cite{ye2025verinabenchmarkingverifiablecode, thakur2025clevercuratedbenchmarkformally}, Coq~\cite{kasibatla2025cobblestonedivideandconquerapproachautomating} or Datalog~\cite{sistla2025verifiedcodereasoningllms}, enabling automated proof checking.
But fully formal methods require formalizing language semantics, which is impractical for arbitrary repository code spanning multiple frameworks and languages.
Sultan et al.~\cite{sultan2026llmsversushaltingproblem} highlight this gap: LLMs can predict program properties like termination with competitive accuracy, yet often fail to provide valid proofs. Moreover, existing approaches tend to be task-specific, requiring separate architectures or training for each problem domain.

We introduce \emph{semi-formal reasoning}, a general approach that bridges this gap.
Rather than training specialized models or formalizing semantics, we prompt agents with structured reasoning templates that require explicit evidence for each claim.
These templates act as \emph{certificates}: the agent must state premises, trace relevant code paths, and provide formal conclusions.
The structured format naturally encourages interprocedural reasoning, as tracing program paths requires the agent to follow function calls rather than guess their behavior.

\paragraph{Motivating Example.}
Figure~\ref{fig:motivating-example} illustrates how these techniques work together on a real patch equivalence task (django\_django-13670). Two patches both attempt to fix 2-digit year formatting for years before 1000 CE:
\begin{center}
\begin{minipage}{0.85\linewidth}
\begin{tcolorbox}[colback=gray!5, colframe=gray!50, boxrule=0.3pt, arc=1pt, left=2pt, right=2pt, top=2pt, bottom=2pt]
\small
\textbf{Patch 1:} \texttt{return format(self.data.year, "04d")[-2:]}\\
\textbf{Patch 2:} \texttt{return '\%02d' \% (self.data.year \% 100)}
\end{tcolorbox}
\end{minipage}
\end{center}
Standard reasoning incorrectly concludes the patches are equivalent, assuming \texttt{format()} is Python's builtin and both produce the same output. With semi-formal analysis, the agent discovers that \texttt{format} is shadowed by a module-level function in Django's \texttt{dateformat.py} that expects a datetime object, not an integer. This causes Patch~1 to raise an \texttt{AttributeError} while Patch~2 succeeds.
Unlike chain-of-thought prompting, which lets the model reason freely, our semi-formal approach requires filling in a structured certificate template with explicit premises, per-test execution traces, and a formal conclusion.
This enforces completeness, ensuring the agent cannot skip cases or make unsupported claims, while remaining in natural language rather than requiring a fully formal proof language.

\begin{figure}[t]
\centering
\begin{tcolorbox}[
  colback=white,
  colframe=gray!70,
  width=\linewidth,
  boxrule=0.5pt,
  arc=3pt,
  title={\textbf{Task:} Are these two patches equivalent (same test outcomes)?},
  fonttitle=\small,
  coltitle=black,
  colbacktitle=gray!15
]
\small

\begin{tcolorbox}[colback=blue!3, colframe=blue!30, boxrule=0.3pt, arc=2pt, left=3pt, right=3pt, top=3pt, bottom=3pt]
\begin{tabular}{@{}p{0.47\linewidth}|p{0.47\linewidth}@{}}
\textbf{Patch 1} & \textbf{Patch 2} \\[0.3em]
\begin{verbatim}
def y(self):
  return format(
    self.data.year, "04d"
  )[-2:]
\end{verbatim}
&
\begin{verbatim}
def y(self):
  return '%02d' % (
    self.data.year % 100
  )
\end{verbatim}
\end{tabular}
\end{tcolorbox}

\vspace{0.4em}

\begin{tcolorbox}[colback=red!5, colframe=red!40, boxrule=0.3pt, arc=2pt, left=3pt, right=3pt, top=2pt, bottom=2pt]
\textbf{Standard Reasoning} \hfill \textcolor{red}{\ding{55} WRONG}\\[0.2em]
\footnotesize ``Both patches produce identical output: \texttt{format(476, '04d')[-2:]} = \texttt{'76'} and \texttt{'\%02d' \% 76} = \texttt{'76'}''
\end{tcolorbox}

\vspace{0.3em}

\begin{tcolorbox}[colback=green!5, colframe=green!50!black, boxrule=0.3pt, arc=2pt, left=3pt, right=3pt, top=2pt, bottom=2pt]
\textbf{Semi-formal (Ours)} \hfill \textcolor{green!50!black}{\ding{51} CORRECT}\\[0.2em]
\footnotesize \emph{Excerpt from the agent's final structured proof:}\\
``\ldots\ \texttt{grep -n 'def format' dateformat.py} $\rightarrow$ line 340: module-level \texttt{format()} shadows Python builtin. This function expects a datetime, not int. \texttt{format(476, '04d')} raises \texttt{AttributeError}. \ldots''
\end{tcolorbox}

\end{tcolorbox}
\caption{Motivating example from Django (django-13670): Standard reasoning assumes \texttt{format()} is Python's builtin; semi-formal analysis traces the actual definition and discovers a module-level function with different semantics, causing Patch~1 to raise \texttt{AttributeError}. See Appendix~\ref{app:mockito8} for a detailed semi-formal proof walkthrough.}
\label{fig:motivating-example}
\end{figure}

We evaluate these techniques on three tasks that test different aspects of code reasoning.
Our primary focus is \emph{patch equivalence verification}: given two patches addressing the same specification, do they produce the same test outcomes? This task provides rigorous ground truth through test execution without requiring expensive human annotation.
We additionally evaluate on \emph{code question answering} (RubberDuckBench) and \emph{fault localization} (Defects4J), which test nuanced semantic understanding and bug finding respectively.
Patch equivalence and fault localization benefit from robust, objective ground truth (test execution outcomes and known buggy lines, respectively), while code question answering relies on expert-written rubrics evaluated by LLM graders.

\paragraph{Contributions.}
\begin{itemize}[noitemsep]
    \item We demonstrate that semi-formal structured reasoning consistently improves agentic code reasoning across three diverse tasks.

    \item For patch equivalence on curated challenging examples, we improve accuracy from 78\% (standard reasoning) to 88\% (semi-formal), a 10 percentage point gain.

    \item For patch equivalence on real-world agent-generated patches with test specifications, we achieve 93\% verification accuracy with Opus-4.5, compared to 86\% for single-shot and 73\% for difflib-based similarity. This enables execution-free feedback for RL training pipelines.

    \item For code question answering on RubberDuckBench~\cite{rubberduckbench}, semi-formal reasoning achieves 87\% accuracy with Opus-4.5, a 10.8 percentage point improvement over single-shot (76\%) and 8.7 percentage points over standard agentic reasoning.

    \item For fault localization on Defects4J~\cite{just2014defects4j}, semi-formal reasoning improves Top-5 accuracy by 5--12 percentage points over standard agentic reasoning.
\end{itemize}

These results demonstrate that LLM agents can perform meaningful semantic code analysis without execution, potentially reducing verification costs in RL training pipelines by avoiding expensive sandbox execution. More broadly, structured agentic reasoning may offer a flexible alternative to classical static analysis tools: rather than encoding analysis logic in specialized algorithms, we can prompt LLM agents with task-specific reasoning templates that generalize across languages and frameworks.

\paragraph{Data Contamination.} SWE-bench instances may appear in LLM training corpora, which could inflate absolute performance numbers. However, our primary conclusions are based on relative comparisons (ablations across reasoning formats) using the same model, where contamination affects all configurations equally. Additionally, our agentic setup requires the model to actively explore repository code at runtime rather than recall memorized solutions.

\section{Background}
\label{sec:background}

We describe our experimental setup and define the three evaluation tasks.

\subsection{Agentic vs. Single-Shot Verification}
We distinguish between single-shot verification, where the model reasons from a static code snapshot, and agentic verification, where the model can actively explore the repository.
Our work focuses on agentic verification using a minimal SWE-agent~\cite{yang2024sweagent} setup with access to bash tools.
We set the maximum number of steps to 100 in all experiments. This allows the verifier to navigate the codebase, trace dependencies, and gather context that would otherwise be unavailable in a single-shot setting.
Crucially, the agent cannot execute repository code or run its test suite: dependencies are not installed and the environment is not sandboxed. The agent may run independent Python scripts to probe general language behavior (e.g., how regex handles edge cases), but its primary mode of analysis is static. Git commands are also disabled to prevent investigating commit history.

\subsection{Patch Equivalence}
True semantic equivalence (determining whether two programs produce identical behavior for all valid inputs) is undecidable in the general case. Real-world code often lacks formal specifications, and developers may hold latent assumptions (e.g., a field \texttt{self.year: int} should only contain positive integers) that LLMs must infer from context. To make this problem tractable, we focus on patch equivalence: given two patches (code diffs that modify a repository) addressing the same task specification, do they produce the same test outcomes? Patches are well-suited for this study because they come with shared specifications (the problem statement) and associated test suites that define expected behavior.

\begin{definition}[Patch Equivalence Modulo Tests]
Two patches are equivalent modulo tests if and only if executing the repository's test suite (F2P $\cup$ P2P) produces identical pass/fail outcomes for both patches.
\end{definition}

\begin{table}[h]
\centering
\caption{Key terminology for patch equivalence}
\label{tab:definitions}
\begin{tabular}{p{3.5cm}p{9cm}}
\toprule
\textbf{Term} & \textbf{Definition} \\
\midrule
Code Patch & A diff that modifies repository source code to fix a bug or add functionality. \\
Test Patch & A diff that adds or modifies test files (the F2P tests introduced with a bug fix). \\
Fail-to-Pass (F2P) Tests & Tests introduced alongside a bug fix that fail before the patch and pass after. These validate that the fix addresses the reported issue. \\
Pass-to-Pass (P2P) / Regression Tests & Existing tests in the repository that must continue passing after the patch is applied. \\
Correct Patch & A patch that causes all F2P tests to pass while not breaking any P2P tests. \\
\bottomrule
\end{tabular}
\end{table}

\subsection{Defects4J for Fault Localization}
Defects4J~\cite{just2014defects4j} is a collection of reproducible bugs from real-world Java projects, widely used for evaluating fault localization and program repair techniques. Each bug includes the buggy and fixed versions of the code, along with at least one failing test that exposes the bug. For fault localization, the task is to identify the exact lines of buggy code given only the failing test, without access to stack traces, error messages, or execution information. Evaluation measures Top-N accuracy: the percentage of bugs where all ground-truth buggy lines are covered within the top N predictions.

\subsection{Code Question Answering}
RubberDuckBench~\cite{rubberduckbench} is a benchmark of 15 code understanding questions across Python, Java, and C++ repositories, each with expert-written rubrics for evaluation. The questions require nuanced understanding of code behavior, including project-specific logic, library semantics, and edge cases. Unlike patch equivalence where ground truth is determined by test execution, code QA requires free-form answers that are evaluated against expert rubrics.

\section{Semi-formal Reasoning}
\label{sec:semiformal}
Given our evaluation tasks, we consider how the verifier should structure its reasoning.
Recent work \cite{achim2025aristotleimolevelautomatedtheorem, olympiaddeepmind, yang2025position} on LLM-based mathematical reasoning has shown that semi-formal approaches, which combine natural language intuition with structured logical steps, often outperform both purely informal and fully formal methods. Fully formal verification in Lean or Coq is tempting but introduces significant overhead: translating arbitrary repository code into a formal representation is nontrivial, and current LLMs still struggle with the precise syntax and proof tactics these systems demand. We therefore compare two reasoning approaches that introduce increasing structure in the agent's reasoning through templates provided in the prompt.

\paragraph{Standard Reasoning}
The agent receives a minimal prompt asking it to determine equivalence, with no structural constraints on its reasoning. It explains its thinking in natural language and concludes with YES/NO. This is fast but allows the agent to make claims without verifying them. For example, an agent might note that one patch modifies extra files but simply assume no tests depend on those changes.

\paragraph{Semi-formal Reasoning}
Semi-formal reasoning adds structure: the agent must state explicit premises, trace execution for each test, and write a formal conclusion. In practice, we enforce this by including a structured template in the agent's initial prompt that specifies the required format.
The key insight is that by structuring the reasoning process, not just the output format, we force the agent to gather evidence before concluding, preventing the premature judgments common in unconstrained reasoning.
We observed that this forces the agent to actually enumerate the program paths rather than make guesses, which naturally leads to deeper interprocedural reasoning as the agent traces function calls to justify its claims.
For instance, in one example (django-13195), semi-formal reasoning caught that session-related tests existed and would fail when one patch omitted changes to the session middleware, something the informal mode missed.
Figure~\ref{fig:motivating-example} illustrates this with a concrete example where tracing a function call reveals a shadowed definition that causes one patch to fail.

The certificate template is task-specific: for patch equivalence, premises describe what each patch modifies and claims trace per-test behavior; for fault localization, premises describe suspicious code regions and claims trace whether each region could cause the observed test failure; for code question answering, the template requires a function trace table, data flow analysis, and semantic properties with explicit evidence. While the specific sections vary by task, all templates enforce the same principle: the agent must document verifiable evidence before reaching a conclusion. Figure~\ref{fig:certificate-template} shows a condensed example for patch equivalence.

\begin{figure}[t]
\centering
\begin{tcolorbox}[
  colback=white,
  colframe=black!60,
  width=\linewidth,
  boxrule=0.5pt,
  arc=2pt,
  title={\textbf{Semi-formal Certificate Template} (patch equivalence)},
  fonttitle=\small,
  coltitle=black,
  colbacktitle=gray!15
]
\small
\begin{verbatim}
DEFINITIONS:
D1: Two patches are EQUIVALENT MODULO TESTS iff
    the test suite produces identical pass/fail
    outcomes for both patches.

PREMISES (state what each patch does):
P1: Patch 1 modifies [file(s)] by [change]
P2: Patch 2 modifies [file(s)] by [change]
P3: The FAIL_TO_PASS tests check [behavior]

ANALYSIS OF TEST BEHAVIOR:
For each test:
  Claim: With Patch 1, test [name] will [PASS/FAIL]
         because [execution trace]
  Claim: With Patch 2, test [name] will [PASS/FAIL]
         because [execution trace]
  Comparison: [SAME/DIFFERENT] outcome

COUNTEREXAMPLE (if NOT EQUIVALENT):
  Test [name] produces different outcomes because
  [specific code trace evidence]

FORMAL CONCLUSION:
By D1, since test outcomes are [IDENTICAL/DIFFERENT],
patches are [EQUIVALENT/NOT EQUIVALENT].
\end{verbatim}
\end{tcolorbox}
\caption{Condensed semi-formal certificate template for patch equivalence. The agent must fill in every bracketed field with evidence gathered from the codebase. The full template includes additional fields for pass-to-pass tests, confidence levels, and alternative hypothesis checks. Templates for other tasks follow the same structure but with task-specific premises and claims. The full prompt appears in Appendix~\ref{app:patch-equiv}.}
\label{fig:certificate-template}
\end{figure}



\section{Evaluation}

We present our experimental results on the three evaluation tasks: patch equivalence verification, code question answering, and fault localization.

\subsection{Patch Equivalence}

\subsubsection{Curated Dataset Evaluation}
We construct a challenging evaluation dataset as follows. A uniformly sampled dataset would be dominated by easily distinguishable patch pairs, offering limited signal for evaluating nuanced reasoning. To better stress-test the verifier, we deliberately curate a more challenging distribution that emphasizes subtle distinctions. We source patches from SWE-bench-Verified, drawing on the community-contributed collection of agent traces, generated patches, and test execution results. From this repository, we sample pairs of patches produced by different agents for the same underlying bug.

Our curation pipeline proceeds as follows. We first score each pair's surface-level similarity using both an LLM-based rating (0 to 10 scale) and difflib-based text similarity. We then exclude pairs with invalid test outcomes. Finally, we balance the dataset to include both high-similarity pairs where patches appear similar (LLM score $\geq$7, difflib $>$0.3) yet differ in resolution status, and positive pairs where both patches pass all tests. The resulting dataset contains 170 challenging examples designed to probe the boundary between superficial code similarity and true semantic equivalence.

Evaluating whether an agent's reasoning is correct step-by-step is difficult as it requires expert review. Hence, in our evaluation we only focus on the final binary answer. Given two patches and a test suite, does the agent correctly predict whether both patches produce the same test outcomes? This is a binary task, and a random baseline achieves 50\% accuracy. We acknowledge that an agent might sometimes arrive at the right answer through flawed reasoning, but the structured formats we introduce next are designed to reduce this by requiring explicit evidence for each claim.

For the experiments in this section the verifier has access to (1) both code patches, (2) the test patch applied to the codebase, and (3) the full repository. The verifier cannot execute tests but can read test implementations.

\begin{table}[h]
\centering
\caption{Comparison of reasoning approaches using Opus-4.5 as the verifier model}
\label{tab:reasoning}
\begin{tabular}{lcccc}
\toprule
\textbf{Reasoning} & \textbf{Overall Accuracy} & \textbf{Avg Steps} & \textbf{Non-Equiv Acc} & \textbf{Equiv Acc} \\
\midrule
Standard & 78.2\% & 10.08 & 78.6\% & 78.0\% \\
Semi-formal & 88.8\% & 28.17 & 82.9\% & 93.0\% \\

\bottomrule
\end{tabular}
\end{table}

Semi-formal reasoning reduces errors by nearly half, improving overall accuracy from 78.2\% to 88.8\% and performing well on both equivalent and non-equivalent cases. This yields a roughly 10-point improvement over informal reasoning but requires about 2.8$\times$ as many steps. Even with semi-formal analysis, 19 cases are incorrect. The primary issue (roughly half) is missed execution differences in non-equivalent cases. The agent claims the patches are equivalent when only one actually passes the tests. This typically happens because the agent fails to trace all relevant code paths or misses some assertions in tests.

\subsubsection{LLM Verifier With Test Specifications}
\label{sec:verifier}

One practical use of patch equivalence is checking whether an LLM-generated patch matches a known reference solution i.e. the gold patch. If we can do this without actually running tests, we get much cheaper feedback for training pipelines. We could avoid the overhead of setting up the execution environment which involves running sandboxes or CI systems for each repository.

The verifier has access to (1) the generated patch, (2) the reference (gold) patch, (3) the test patch applied to the codebase (i.e., F2P test code is visible), and (4) the full repository including all P2P tests. Ground truth is determined by actual test execution.

We tested this by comparing patches from live-swe-agent (gemini-3-pro) against the correct patches from SWE-bench-Verified dataset. Unlike the curated dataset in the previous section, here we sample randomly: we selected a balanced dataset of 200 examples, evenly split between correct and incorrect patches (100 each). The verifier's task is to check if the agent's patch is equivalent to the reference solution. We evaluate the following verification modes:
\begin{itemize}
    \item \textbf{difflib (No LLM):} Recent work from SWE-RL~\cite{wei2025swerl} used the Python difflib library to compute a similarity score as the RL reward. For binary classification, one could set a threshold on this score, above which patches are considered equivalent. We evaluated various threshold values on the test set and found that a threshold of 0.4 yields the maximum accuracy. However, note that this optimal threshold may not be known a priori for a given benchmark. We present this value here in the table for reference.

    \item \textbf{Single LLM Call:} One LLM call with just the problem statement and patches. No tool use or codebase exploration.

    \item \textbf{Single LLM Call + File Context:} Same as above, but the prompt includes the full contents of all files modified by the patches.

    \item \textbf{Agentic:} The verifier explores the codebase using tools over multiple steps, reasoning from the problem description and existing code. We evaluate both standard reasoning and semi-formal reasoning formats.
\end{itemize}
In all settings, accuracy is computed against actual test execution results.
\begin{table}[h]
\centering
\caption{Verification accuracy with test patches available}
\label{tab:verifier-full}
\begin{tabular}{llcc}
\toprule
\textbf{Model} & \textbf{Mode} & \textbf{Accuracy} & \textbf{Avg Steps} \\
\midrule
difflib & Best threshold=0.4 & 73\% & - \\
\midrule
Sonnet-4.5 & Single Call (Standard) & 80.0\% & 1 \\
Sonnet-4.5 & + File Context (Standard) & 82.0\% & 1 \\
Sonnet-4.5 & Agentic (Standard) & 84.5\% & 20.67 \\
Sonnet-4.5 & Agentic (Semi-formal) & \textbf{91.5\%} & 43.23 \\
\midrule
Opus-4.5 & Single Call (Standard) & 86.0\% & 1 \\
Opus-4.5 & + File Context (Standard) & 87.5\% & 1 \\
Opus-4.5 & Agentic (Standard) & 87.0\% & 19.7 \\
Opus-4.5 & Agentic (Semi-formal) & \textbf{93.0\%} & 37.82 \\
\bottomrule
\end{tabular}
\end{table}
Opus-4.5 achieves 93.0\% accuracy with semi-formal reasoning when given the fail-to-pass test patch. The single-call modes (80--87\% accuracy) confirm that many verification tasks require exploring code beyond the immediate patch context.

\textbf{Error Analysis for Verifier With Tests.}
We closely analyzed the remaining errors for Opus-4.5 with Semi-formal and found three primary failure modes:
\begin{enumerate}
    \item \textbf{Incomplete execution tracing despite the given instructions:} The agent assumed function behavior without fully tracing concrete execution paths.

    \item \textbf{Third-party library semantics:} The agent guessed behavior from function names when source code was unavailable.

    \item \textbf{Dismissing subtle differences:} The agent identified semantic differences but incorrectly concluded they were irrelevant to test outcomes.
\end{enumerate}
While these results demonstrate strong verification performance when fail-to-pass test patches are available, this dependency limits applicability to scenarios without well-defined test patches. In the following section, we explore extending our approach to scenarios where test patches are unavailable.

\subsection{Fault Localization}

We evaluate fault localization on Defects4J (see Section~\ref{sec:background}). Unlike patch equivalence where both patches are provided, fault localization requires the agent to \emph{find} the relevant code in a large search space, then reason about why it causes the test failure.
Appendix~\ref{app:fl-prompt} shows the prompt we use for structured reasoning.

\paragraph{Metric}
For matching, we compare overlaps against buggy file regions: the lines in the patch, grouped per hunk into (file, min\_deleted, max\_deleted) ranges for deletions, and (file, insert\_pos, insert\_pos) for insertions. A prediction matches a region if their ranges overlap: pred\_start $\leq$ region\_end and pred\_end $\geq$ region\_start. This works for all patch types including deletion-only cases, which can be an issue with matching based on line number in the ground truth patch. We report two variants: \textbf{All}, where a bug is solved at Top-N only when every ground-truth hunk is covered by predictions in positions 1..N, and \textbf{Any}, where at least one ground-truth hunk must be covered. The All metric is stricter and penalizes multi-hunk bugs; the Any metric better reflects partial localization success.

\paragraph{Available Information.} The agent has access to (1) the failing test name and source code, (2) source files scoped to classes loaded during test execution, and (3) the full repository. No stack traces, error messages, or execution information are provided (to avoid trivializing the task).

\paragraph{Small-Scale Evaluation (50 bugs).} We first evaluate on 50 bugs from Defects4J where all relevant source files fit within the context window, enabling comparison between single-shot (all code provided upfront) and agentic (iterative file exploration) modes.  We do this by estimating the token count in loaded classes and cap it to 100K.
These projects contained 43 unique bugs that were evaluable; others had missing source or test files, or other errors.

\begin{table}[h]
\centering
\caption{Fault localization accuracy on Defects4J (Opus-4.5). ``All'' requires every ground-truth hunk to be covered; ``Any'' requires at least one.}
\label{tab:fl-results}
\begin{tabular}{llcccccc}
\toprule
 & & \multicolumn{3}{c}{\textbf{All}} & \multicolumn{3}{c}{\textbf{Any}} \\
\cmidrule(lr){3-5} \cmidrule(lr){6-8}
\textbf{Mode} & \textbf{Exploration} & \textbf{Top-1} & \textbf{Top-3} & \textbf{Top-5} & \textbf{Top-1} & \textbf{Top-3} & \textbf{Top-5} \\
\midrule
Standard & Single-shot & 36.1\% & 55.6\% & 55.6\% & 47.2\% & 69.4\% & 69.4\% \\
Semi-formal & Single-shot & 41.7\% & 58.3\% & 63.9\% & 55.6\% & 75.0\% & 77.8\% \\
Standard & Agentic & 46.5\% & 60.5\% & 60.5\% & 65.1\% & 79.1\% & 81.4\% \\
Semi-formal & Agentic & \textbf{53.5\%} & \textbf{67.4\%} & \textbf{72.1\%} & \textbf{72.1\%} & \textbf{83.7\%} & \textbf{88.4\%} \\
\bottomrule
\end{tabular}
\end{table}

Semi-formal reasoning improves accuracy across both metrics and exploration modes. Under the stricter All metric, Top-5 accuracy improves by +8pp for single-shot and +12pp for agentic. Under the Any metric, the gains are similar: +8pp for single-shot and +7pp for agentic. Agentic exploration with semi-formal reasoning achieves the best overall accuracy at 72.1\% Top-5 (All) and 88.4\% Top-5 (Any).

\paragraph{Larger-Scale Validation (100 bugs).}
To validate our findings on a larger sample, we evaluated on 100 Defects4J bugs randomly sampled across 14 Java projects (of which 90 were evaluable). Unlike the small-scale study, many of these bugs involve source files exceeding context limits, requiring genuine agentic exploration. Standard mode sometimes makes a quick, correct prediction on turn 1 from just the test code (e.g., Closure\_129, where the agent predicts correctly from the test code alone without reading any source files).

Table~\ref{tab:fl-58} shows semi-formal reasoning improves accuracy by +5pp Top-5 (All) over standard, confirming that semi-formal reasoning consistently outperforms standard mode across dataset sizes. The Any metric, which requires at least one ground-truth hunk to be covered, shows the same trend with higher absolute numbers.

\begin{table}[h]
\centering
\caption{Agentic fault localization on 100-bug sample. ``All'' requires every ground-truth hunk to be covered; ``Any'' requires at least one.}
\label{tab:fl-58}
\begin{tabular}{llcccccc}
\toprule
 & & \multicolumn{3}{c}{\textbf{All}} & \multicolumn{3}{c}{\textbf{Any}} \\
\cmidrule(lr){3-5} \cmidrule(lr){6-8}
\textbf{Model} & \textbf{Mode} & \textbf{Top-1} & \textbf{Top-3} & \textbf{Top-5} & \textbf{Top-1} & \textbf{Top-3} & \textbf{Top-5} \\
\midrule
Sonnet-4.5 & Standard & 20.0\% & 30.0\% & 31.1\% & 35.6\% & 51.1\% & 51.1\% \\
Sonnet-4.5 & Semi-formal & 16.7\% & 25.6\% & 30.0\% & 30.0\% & 43.3\% & 50.0\% \\
Opus-4.5 & Standard & 30.0\% & 42.2\% & 43.3\% & 53.3\% & 64.4\% & 65.6\% \\
Opus-4.5 & Semi-formal & \textbf{34.4\%} & \textbf{47.8\%} & \textbf{47.8\%} & \textbf{58.9\%} & \textbf{68.9\%} & \textbf{68.9\%} \\
\bottomrule
\end{tabular}
\end{table}

\paragraph{Error Analysis.}
We analyzed failures for Opus-4.5 with semi-formal reasoning and identified four primary failure modes:
\begin{enumerate}
    \item \textbf{Indirection bugs:} The bug is in a class not directly invoked by the test. For example, in Csv\_12, the test calls \texttt{CSVParser.parse()} but the bug is in \texttt{CSVFormat.withHeader()}---a configuration class the model consistently overlooks.

    \item \textbf{Multi-file bugs:} Bugs spanning multiple files require identifying all locations. Large ground-truth sets (5+ lines across 2+ files) are systematically harder.

    \item \textbf{Domain-specific bugs:} Algorithmic bugs requiring specialized knowledge, such as numerical analysis issues in Math\_81 (EigenDecomposition), exceed the model's domain expertise.

    \item \textbf{More than 5 fix regions:}
    In a few cases (5/90), the number of distinct regions of changes in the ground truth was more than 5, which will result in a miss by our metric.
\end{enumerate}

Appendix~\ref{app:mockito8} gives a walkthrough of how semi-formal reasoning helps in the case of mockito\_8.






\subsection{Code Question Answering}

We evaluate code question answering on RubberDuckBench (see Section~\ref{sec:background}).

\paragraph{Available Information.} In single-shot mode, the agent receives only the function containing the code in question (approximately 20--50 lines of context). In agentic mode, the agent can explore the full repository using bash tools to read files, search for definitions, and trace dependencies.

\paragraph{Evaluation.} Since code QA requires nuanced assessment, we use multi-model LLM grading with rubric-based evaluation. Each answer is graded independently by Gemini-3-Pro and GPT-5.2 against the expert rubric, with weighted averaging to handle disagreements. The graders achieved 85\% agreement across all evaluations, indicating consistent assessment.

\paragraph{Reasoning Modes.} We evaluate two reasoning approaches:
\begin{itemize}[noitemsep]
    \item \textbf{Standard}: Simple prompt asking for a code-grounded answer.
    \item \textbf{Semi-formal}:  Structured output template requiring function trace tables, data flow analysis, and explicit evidence citations.
\end{itemize}

\begin{table}[h]
\centering
\caption{Code question answering accuracy on RubberDuckBench}
\label{tab:rubber-duck}
\begin{tabular}{llcc}
\toprule
\textbf{Model} & \textbf{Mode} & \textbf{Accuracy} & \textbf{Avg Steps} \\
\midrule
Opus-4.5 & Single-shot & 76.2\% & 1 \\
Opus-4.5 & Agentic (Standard) & 78.3\% & 10.8 \\
Opus-4.5 & Agentic (Semi-formal) & \textbf{87.0\%} & 19.7 \\
\midrule
Sonnet-4.5 & Single-shot & 78.7\% & 1 \\
Sonnet-4.5 & Agentic (Standard) & 84.2\% & 17.6 \\
Sonnet-4.5 & Agentic (Semi-formal) & 84.8\% & 25.3 \\
\bottomrule
\end{tabular}
\end{table}

The results reveal that \emph{structured semi-formal reasoning provides substantial gains}: Opus improves from 78.3\% (standard agentic) to 87.0\% with the semi-formal template (+8.7pp). For Sonnet, standard agentic reasoning already achieves 85.3\%, and the semi-formal template does not yield further gains (84.8\%), suggesting that the benefit of structured reasoning varies by model capability and may plateau when the base model is already strong. For Opus, the structured template forces the agent to document evidence systematically before answering, correcting its tendency to guess based on function names.

The structured template requires the agent to fill in a function trace table (listing every function examined with file:line locations and verified behavior), data flow analysis (tracing how key variables flow through the code), semantic properties with explicit evidence, and an alternative hypothesis check. This structured format reduces the tendency to guess based on function names, a common failure mode we observed in unstructured reasoning.

\paragraph{Error Analysis.}
The structured template forces the agent to document evidence systematically before reaching conclusions. For example, on a question about whether two API calls differ (cpp\_3), standard reasoning stated ``it provides proper error handling if an invalid key is somehow passed''---implying edge cases could occur. The semi-formal template required tracing both the map initialization and variable assignments, revealing they use the same enum values, thus proving invalid keys are impossible. This explicit verification step eliminated a deduction that standard reasoning incurred. Conversely, semi-formal reasoning can fail when agents construct elaborate but incomplete reasoning chains: on py\_5, the agent thoroughly traced five functions but missed that downstream code already handled the edge case it identified, leading to a confident but wrong answer. See Appendix~\ref{app:rubber-duck} for detailed examples with full reasoning traces.

\section{Related Work}

\paragraph{LLM-Based Software Engineering Agents.}
The emergence of LLM-based coding agents has transformed automated software engineering.
SWE-agent~\cite{yang2024sweagent} introduced an agent-computer interface that enables LLMs to interact with codebases through specialized commands, achieving strong results on SWE-bench~\cite{jimenez2024swebench}.
OpenHands~\cite{wang2024openhands} provides an open platform for building software development agents.
Agentless~\cite{xia2024agentless} takes a different approach, decomposing bug fixing into localization and repair phases without persistent agent state.
These systems rely on test execution for validation, which our work aims to reduce through semantic code reasoning.

\paragraph{Execution-Free Verification for Code.}
Most closely related to our work are recent approaches to verifying code without execution.
SWE-RM~\cite{shum2025swerm} trains a reward model to provide execution-free feedback for software engineering agents, showing that learned verifiers can approximate test outcomes.
Agentic Rubrics~\cite{raghavendra2025rubrics} propose using LLM-generated rubrics as contextual verifiers, decomposing verification into interpretable criteria.
CodeJudge~\cite{tong2024codejudge} explores LLM-as-a-judge for evaluating generated code quality.
Our approach differs by emphasizing structured semi-formal reasoning to improve verification accuracy, achieving 93\% accuracy on real-world patches.

\paragraph{LLM-Based Fault Localization and Code Understanding.}
Beyond patch verification, agentic code reasoning encompasses fault localization and code understanding.
AgentFL~\cite{qin2024agentfl} uses LLM agents for project-level fault localization, while FlexFL~\cite{xu2025flexflflexibleeffectivefault} demonstrates effective fault localization with open-source LLMs.
For repository-level code understanding, CodePlan~\cite{bairi2024codeplan} combines LLMs with planning for multi-step repository edits.
RubberDuckBench~\cite{rubberduckbench} provides a benchmark for evaluating AI coding assistants on code understanding tasks, measuring how well LLMs can answer questions about codebases.
These works highlight the broader applicability of agentic reasoning for code analysis tasks beyond verification.

\paragraph{Program Equivalence and Formal Verification.}
Program equivalence is a fundamental problem in computer science, known to be undecidable in the general case~\cite{rice1953classes}.
Traditional approaches rely on formal methods such as translation validation~\cite{pnueli1998translation} and equivalence checking~\cite{necula2000translation}.
Recent work has explored using LLMs for formal verification~\cite{first2023baldur}, though translating arbitrary code to formal specifications remains challenging.
EquiBench~\cite{wei-etal-2025-equibench} benchmarks LLM reasoning about program equivalence across several transformation types, but focuses on small self-contained code pairs rather than repository-level patches with test suites.
Sultan et al.~\cite{sultan2026llmsversushaltingproblem} show that LLMs can predict program termination with surprising accuracy, ranking competitively with specialized tools on SV-Comp benchmarks, though they often fail to provide valid proofs.

A complementary line of work focuses on post-hoc verification of LLM reasoning~\cite{sistla2025verifiedcodereasoningllms}, translating LLM responses into Datalog facts and using static analysis to verify reasoning steps, successfully validating judgments on uninitialized variable errors and catching incorrect equivalence judgments.
Our approach differs in focusing on the ``input side'': improving what the agent is asked to do through structured semi-formal reasoning, rather than post-facto output verification. The semi-formal certificates we require are designed to be easier to manually validate than examining full agent trajectories, though they lack the automated checkability of fully formal approaches. These approaches are complementary: structured reasoning improves agent thoroughness during analysis, while formal verification can provide additional guarantees on outputs.

\paragraph{LLM Reasoning and Chain-of-Thought.}
Our semi-formal reasoning approach builds on work showing that structured reasoning improves LLM performance.
Chain-of-thought prompting~\cite{wei2022chain} demonstrated that intermediate reasoning steps improve mathematical problem solving. ReAct~\cite{yao2023react} combines reasoning with action for agent tasks. CodeAct~\cite{wang2024codeact} shows that executable code actions improve agent performance.
We extend these ideas to code reasoning, showing that task-specific structured templates (requiring premises, execution traces, and formal conclusions) improve semantic verification accuracy by up to 11 percentage points.

\paragraph{Training and Scaling SWE Agents.}
Recent work has focused on training pipelines for software engineering agents.
SWE-Gym~\cite{pan2024swegym} provides a training environment with 2,438 real-world Python tasks, enabling both agent and verifier training through reinforcement learning.
R2E-Gym~\cite{jain2025r2egym} scales this further with procedural environment generation and hybrid verifiers combining execution-based and LLM-based feedback, training agents on 89K instances.
SWE-RL~\cite{wei2025swerl} advances reasoning capabilities through reinforcement learning on open software evolution, using difflib-based similarity as a reward signal.
Our work complements these efforts by providing execution-free verification that could reduce the computational cost of RL training.

\section{Conclusion and Future Work}
We studied agentic code reasoning across three tasks: patch equivalence verification, code question answering, and fault localization. Our key findings include:
\begin{itemize}
    \item Semi-formal structured reasoning consistently improves agentic code reasoning across all three tasks, with gains of 5--12 percentage points over standard agentic baselines.

    \item For patch equivalence, we achieve 93\% verification accuracy on real-world patches, a 7 percentage point improvement over single-shot baselines, enabling execution-free feedback for RL training pipelines.

    \item For code question answering, semi-formal reasoning achieves 87\% accuracy on RubberDuckBench, a 9 percentage point gain over standard agentic reasoning.

    \item For fault localization on Defects4J, semi-formal reasoning consistently improves accuracy over standard reasoning, with gains of up to 12 percentage points on fit-in-context bugs and 5 percentage points on a larger 90-bug evaluation.
\end{itemize}
These results demonstrate that LLM agents can perform meaningful semantic code analysis without execution. Structured reasoning templates offer a complementary approach to classical static analysis: rather than encoding analysis logic in specialized algorithms, we can prompt agents with task-specific formats that generalize across languages and frameworks, though without the formal guarantees of traditional tools.

Several directions for future work emerge:
\begin{itemize}


    \item \textbf{Post-training for code reasoning:} Fine-tuning models to internalize the semi-formal template structure could further improve accuracy while potentially eliminating the prompt overhead.

    \item \textbf{Extending to other static analysis tasks:} The semi-formal reasoning approach could be applied to other code analysis tasks such as security vulnerability detection, code smell identification, and API misuse detection.

    \item \textbf{Hybrid verification:} Combining LLM-based reasoning with lightweight formal methods or symbolic execution could provide stronger guarantees while maintaining flexibility.
\end{itemize}

\bibliographystyle{ACM-Reference-Format}
\bibliography{ref}

@misc{kasibatla2025cobblestonedivideandconquerapproachautomating,
      title={Cobblestone: A Divide-and-Conquer Approach for Automating Formal Verification}, 
      author={Saketh Ram Kasibatla and Arpan Agarwal and Yuriy Brun and Sorin Lerner and Talia Ringer and Emily First},
      year={2025},
      eprint={2410.19940},
      archivePrefix={arXiv},
      primaryClass={cs.LO},
      url={https://arxiv.org/abs/2410.19940}, 
}

@misc{ye2025verinabenchmarkingverifiablecode,
      title={VERINA: Benchmarking Verifiable Code Generation}, 
      author={Zhe Ye and Zhengxu Yan and Jingxuan He and Timothe Kasriel and Kaiyu Yang and Dawn Song},
      year={2025},
      eprint={2505.23135},
      archivePrefix={arXiv},
      primaryClass={cs.LG},
      url={https://arxiv.org/abs/2505.23135}, 
}

@misc{thakur2025clevercuratedbenchmarkformally,
      title={CLEVER: A Curated Benchmark for Formally Verified Code Generation}, 
      author={Amitayush Thakur and Jasper Lee and George Tsoukalas and Meghana Sistla and Matthew Zhao and Stefan Zetzsche and Greg Durrett and Yisong Yue and Swarat Chaudhuri},
      year={2025},
      eprint={2505.13938},
      archivePrefix={arXiv},
      primaryClass={cs.LG},
      url={https://arxiv.org/abs/2505.13938}, 
}

@misc{achim2025aristotleimolevelautomatedtheorem,
      title={Aristotle: IMO-level Automated Theorem Proving}, 
      author={Tudor Achim and Alex Best and Alberto Bietti and Kevin Der and Mathïs Fédérico and Sergei Gukov and Daniel Halpern-Leistner and Kirsten Henningsgard and Yury Kudryashov and Alexander Meiburg and Martin Michelsen and Riley Patterson and Eric Rodriguez and Laura Scharff and Vikram Shanker and Vladmir Sicca and Hari Sowrirajan and Aidan Swope and Matyas Tamas and Vlad Tenev and Jonathan Thomm and Harold Williams and Lawrence Wu},
      year={2025},
      eprint={2510.01346},
      archivePrefix={arXiv},
      primaryClass={cs.AI},
      url={https://arxiv.org/abs/2510.01346}, 
}

@article{olympiaddeepmind,
title	= {Olympiad-Level Formal Mathematical Reasoning with Reinforcement Learning},
author	= {Thomas Hubert and Rishi Mehta and Laurent Sartran and Miklós Z. Horváth and Goran Žužić and Eric Wieser and Aja Huang and Julian Schrittwieser and Yannick Schroecker and Hussain Masoom and Ottavia Bertolli and Tom Zahavy and Amol Mandhane and Jessica Yung and Iuliya Beloshapka and Borja Ibarz and Vivek Veeriah and Lei Yu and Oliver Nash and Paul Lezeau and Salvatore Mercuri and Calle Sönne and Bhavik Mehta and Alex Davies and Daniel Zheng and Fabian Pedregosa and Yin Li and Ingrid von Glehn and Mark Rowland and Samuel Albanie and Ameya Velingker and Simon Schmitt and Edward Lockhart and Henryk Michalewski and Nicolas Sonnerat and Demis Hassabis and Pushmeet Kohli and David Silver},year	= {2025},URL	= {https://www.nature.com/articles/s41586-025-09833-y},journal	= {Nature}}

@inproceedings{
yang2025position,
title={Position: Formal Mathematical Reasoning{\textemdash}A New Frontier in {AI}},
author={Kaiyu Yang and Gabriel Poesia and Jingxuan He and Wenda Li and Kristin E. Lauter and Swarat Chaudhuri and Dawn Song},
booktitle={Forty-second International Conference on Machine Learning Position Paper Track},
year={2025},
url={https://openreview.net/forum?id=HuvAM5x2xG}
}

@inproceedings{just2014defects4j,
  author = {Just, Ren{\'e} and Jalali, Darioush and Ernst, Michael D.},
  title = {Defects4J: A Database of Existing Faults to Enable Controlled Testing Studies for Java Programs},
  year = {2014},
  isbn = {9781450330138},
  publisher = {Association for Computing Machinery},
  address = {New York, NY, USA},
  url = {https://doi.org/10.1145/2610384.2628055},
  doi = {10.1145/2610384.2628055},
  booktitle = {Proceedings of the 2014 International Symposium on Software Testing and Analysis},
  pages = {437--440},
  numpages = {4},
  location = {San Jose, CA, USA},
  series = {ISSTA 2014}
}

@misc{sistla2025verifiedcodereasoningllms,
      title={Towards Verified Code Reasoning by LLMs},
      author={Meghana Sistla and Gogul Balakrishnan and Pat Rondon and José Cambronero and Michele Tufano and Satish Chandra},
      year={2025},
      eprint={2509.26546},
      archivePrefix={arXiv},
      primaryClass={cs.SE},
      url={https://arxiv.org/abs/2509.26546},
}

@inproceedings{
yang2024sweagent,
title={{SWE}-agent: Agent-Computer Interfaces Enable Automated Software Engineering},
author={John Yang and Carlos E Jimenez and Alexander Wettig and Kilian Lieret and Shunyu Yao and Karthik R Narasimhan and Ofir Press},
booktitle={The Thirty-eighth Annual Conference on Neural Information Processing Systems},
year={2024},
url={https://openreview.net/forum?id=mXpq6ut8J3}
}

@inproceedings{
jimenez2024swebench,
title={{SWE}-bench: Can Language Models Resolve Real-world Github Issues?},
author={Carlos E Jimenez and John Yang and Alexander Wettig and Shunyu Yao and Kexin Pei and Ofir Press and Karthik R Narasimhan},
booktitle={The Twelfth International Conference on Learning Representations},
year={2024},
url={https://openreview.net/forum?id=VTF8yNQM66}
}

@inproceedings{
wang2024openhands,
title={OpenHands: An Open Platform for {AI} Software Developers as Generalist Agents},
author={Xingyao Wang and Boxuan Li and Yufan Song and Frank F. Xu and Xiangru Tang and Mingchen Zhuge and Jiayi Pan and Yueqi Song and Bowen Li and Jaskirat Singh and Hoang H. Tran and Fuqiang Li and Ren Ma and Mingzhang Zheng and Bill Qian and Yanjun Shao and Niklas Muennighoff and Yizhe Zhang and Binyuan Hui and Junyang Lin and Robert Brennan and Hao Peng and Heng Ji and Graham Neubig},
booktitle={The Thirteenth International Conference on Learning Representations},
year={2025},
url={https://openreview.net/forum?id=OJd3ayDDoF}
}

@misc{xia2024agentless,
      title={Agentless: Demystifying LLM-based Software Engineering Agents},
      author={Chunqiu Steven Xia and Yinlin Deng and Soren Dunn and Lingming Zhang},
      year={2024},
      eprint={2407.01489},
      archivePrefix={arXiv},
      primaryClass={cs.SE},
      url={https://arxiv.org/abs/2407.01489},
}

@misc{pan2024swegym,
      title={Training Software Engineering Agents and Verifiers with SWE-Gym},
      author={Jiayi Pan and Xingyao Wang and Graham Neubig and Navdeep Jaitly and Heng Ji and Alane Suhr and Yizhe Zhang},
      year={2025},
      eprint={2412.21139},
      archivePrefix={arXiv},
      primaryClass={cs.SE},
      url={https://arxiv.org/abs/2412.21139},
}

@misc{jain2025r2egym,
      title={R2E-Gym: Procedural Environments and Hybrid Verifiers for Scaling Open-Weights SWE Agents},
      author={Naman Jain and Jaskirat Singh and Manish Shetty and Liang Zheng and Koushik Sen and Ion Stoica},
      year={2025},
      eprint={2504.07164},
      archivePrefix={arXiv},
      primaryClass={cs.SE},
      url={https://arxiv.org/abs/2504.07164},
}

@inproceedings{
wei2025swerl,
title={{SWE}-{RL}: Advancing {LLM} Reasoning via Reinforcement Learning on Open Software Evolution},
author={Yuxiang Wei and Olivier Duchenne and Jade Copet and Quentin Carbonneaux and LINGMING ZHANG and Daniel Fried and Gabriel Synnaeve and Rishabh Singh and Sida Wang},
booktitle={The Thirty-ninth Annual Conference on Neural Information Processing Systems},
year={2025},
url={https://openreview.net/forum?id=ULblO61XZ0}
}

@misc{shum2025swerm,
      title={SWE-RM: Execution-free Feedback For Software Engineering Agents},
      author={KaShun Shum and Binyuan Hui and Jiawei Chen and Lei Zhang and X. W. and Jiaxi Yang and Yuzhen Huang and Junyang Lin and Junxian He},
      year={2025},
      eprint={2512.21919},
      archivePrefix={arXiv},
      primaryClass={cs.CL},
      url={https://arxiv.org/abs/2512.21919},
}

@misc{raghavendra2025rubrics,
      title={Agentic Rubrics as Contextual Verifiers for SWE Agents},
      author={Mohit Raghavendra and Anisha Gunjal and Bing Liu and Yunzhong He},
      year={2026},
      eprint={2601.04171},
      archivePrefix={arXiv},
      primaryClass={cs.LG},
      url={https://arxiv.org/abs/2601.04171},
}

@misc{tong2024codejudge,
      title={CodeJudge: Evaluating Code Generation with Large Language Models},
      author={Weixi Tong and Tianyi Zhang},
      year={2024},
      eprint={2410.02184},
      archivePrefix={arXiv},
      primaryClass={cs.LG},
      url={https://arxiv.org/abs/2410.02184},
}

@misc{qin2024agentfl,
      title={AgentFL: Scaling LLM-based Fault Localization to Project-Level Context},
      author={Yihao Qin and Shangwen Wang and Yiling Lou and Jinhao Dong and Kaixin Wang and Xiaoling Li and Xiaoguang Mao},
      year={2025},
      eprint={2403.16362},
      archivePrefix={arXiv},
      primaryClass={cs.SE},
      url={https://arxiv.org/abs/2403.16362},
}

@misc{xu2025flexflflexibleeffectivefault,
      title={FlexFL: Flexible and Effective Fault Localization with Open-Source Large Language Models},
      author={Chuyang Xu and Zhongxin Liu and Xiaoxue Ren and Gehao Zhang and Ming Liang and David Lo},
      year={2025},
      eprint={2411.10714},
      archivePrefix={arXiv},
      primaryClass={cs.SE},
      url={https://arxiv.org/abs/2411.10714},
}

@misc{bairi2024codeplan,
      title={CodePlan: Repository-level Coding using LLMs and Planning},
      author={Ramakrishna Bairi and Atharv Sonwane and Aditya Kanade and Vageesh D C and Arun Iyer and Suresh Parthasarathy and Sriram Rajamani and B. Ashok and Shashank Shet},
      year={2023},
      eprint={2309.12499},
      archivePrefix={arXiv},
      primaryClass={cs.SE},
      url={https://arxiv.org/abs/2309.12499},
}

@misc{rubberduckbench,
  title={{RubberDuckBench}: A Benchmark for {AI} Coding Assistants},
  author={Mohammad, Ferida and Ayad, Fatma and Maniatis, Petros and Chandra, Satish and Dinella, Elizabeth},
  journal={arXiv preprint arXiv:2601.16456},
  year={2026}
}

@article{rice1953classes,
  author = {Rice, H. G.},
  journal = {Transactions of the American Mathematical Society},
  number = 2,
  pages = {358--366},
  publisher = {American Mathematical Society},
  title = {Classes of recursively enumerable sets and their decision problems},
  volume = 74,
  year = 1953
}

@inproceedings{pnueli1998translation,
author = {Pnueli, Amir and Siegel, Michael and Singerman, Eli},
title = {Translation Validation},
year = {1998},
isbn = {3540643567},
publisher = {Springer-Verlag},
address = {Berlin, Heidelberg},
booktitle = {Proceedings of the 4th International Conference on Tools and Algorithms for Construction and Analysis of Systems},
pages = {151–166},
numpages = {16},
series = {TACAS '98}
}

@inproceedings{necula2000translation,
author = {Necula, George C.},
title = {Translation validation for an optimizing compiler},
year = {2000},
isbn = {1581131992},
publisher = {Association for Computing Machinery},
address = {New York, NY, USA},
url = {https://doi.org/10.1145/349299.349314},
doi = {10.1145/349299.349314},
booktitle = {Proceedings of the ACM SIGPLAN 2000 Conference on Programming Language Design and Implementation},
pages = {83–94},
numpages = {12},
location = {Vancouver, British Columbia, Canada},
series = {PLDI '00}
}

@misc{first2023baldur,
      title={Baldur: Whole-Proof Generation and Repair with Large Language Models},
      author={Emily First and Markus N. Rabe and Talia Ringer and Yuriy Brun},
      year={2023},
      eprint={2303.04910},
      archivePrefix={arXiv},
      primaryClass={cs.LG},
      url={https://arxiv.org/abs/2303.04910},
}

@inproceedings{wei2022chain,
author = {Wei, Jason and Wang, Xuezhi and Schuurmans, Dale and Bosma, Maarten and Ichter, Brian and Xia, Fei and Chi, Ed H. and Le, Quoc V. and Zhou, Denny},
title = {Chain-of-thought prompting elicits reasoning in large language models},
year = {2022},
isbn = {9781713871088},
publisher = {Curran Associates Inc.},
address = {Red Hook, NY, USA},
booktitle = {Proceedings of the 36th International Conference on Neural Information Processing Systems},
articleno = {1800},
numpages = {14},
location = {New Orleans, LA, USA},
series = {NIPS '22}
}

@misc{yao2023react,
      title={ReAct: Synergizing Reasoning and Acting in Language Models},
      author={Shunyu Yao and Jeffrey Zhao and Dian Yu and Nan Du and Izhak Shafran and Karthik Narasimhan and Yuan Cao},
      year={2023},
      eprint={2210.03629},
      archivePrefix={arXiv},
      primaryClass={cs.CL},
      url={https://arxiv.org/abs/2210.03629},
}

@misc{wang2024codeact,
      title={Executable Code Actions Elicit Better LLM Agents},
      author={Xingyao Wang and Yangyi Chen and Lifan Yuan and Yizhe Zhang and Yunzhu Li and Hao Peng and Heng Ji},
      year={2024},
      eprint={2402.01030},
      archivePrefix={arXiv},
      primaryClass={cs.CL},
      url={https://arxiv.org/abs/2402.01030},
}

@misc{sultan2026llmsversushaltingproblem,
      title={LLMs versus the Halting Problem: Revisiting Program Termination Prediction},
      author={Oren Sultan and Jordi Armengol-Estape and Pascal Kesseli and Julien Vanegue and Dafna Shahaf and Yossi Adi and Peter O'Hearn},
      year={2026},
      eprint={2601.18987},
      archivePrefix={arXiv},
      primaryClass={cs.CL},
      url={https://arxiv.org/abs/2601.18987},
}

@inproceedings{wei-etal-2025-equibench,
    title = "{E}qui{B}ench: Benchmarking Large Language Models' Reasoning about Program Semantics via Equivalence Checking",
    author = "Wei, Anjiang  and
      Cao, Jiannan  and
      Li, Ran  and
      Chen, Hongyu  and
      Zhang, Yuhui  and
      Wang, Ziheng  and
      Liu, Yuan  and
      Teixeira, Thiago S. F. X.  and
      Yang, Diyi  and
      Wang, Ke  and
      Aiken, Alex",
    editor = "Christodoulopoulos, Christos  and
      Chakraborty, Tanmoy  and
      Rose, Carolyn  and
      Peng, Violet",
    booktitle = "Proceedings of the 2025 Conference on Empirical Methods in Natural Language Processing",
    month = nov,
    year = "2025",
    address = "Suzhou, China",
    publisher = "Association for Computational Linguistics",
    url = "https://aclanthology.org/2025.emnlp-main.1718/",
    doi = "10.18653/v1/2025.emnlp-main.1718",
    pages = "33868--33881",
    ISBN = "979-8-89176-332-6"
}

\appendix
\section{Patch Equivalence Semi-formal Reasoning Template}
\label{app:patch-equiv}

\paragraph{Semi-formal Prompt Template.}
The structured template used for semi-formal reasoning in patch equivalence verification requires the agent to construct a formal proof of equivalence or non-equivalence:

\begin{tcolorbox}[
  title=Semi-formal Proof of Patch Equivalence,
  colback=gray!5,
  colframe=gray!75!black,
  fonttitle=\bfseries,
  breakable,
  enhanced
]
\begin{verbatim}
DEFINITIONS:
D1: Two patches are EQUIVALENT MODULO TESTS iff executing the
    existing repository test suite produces identical pass/fail
    outcomes for both patches.
D2: The relevant tests are ONLY those in FAIL_TO_PASS and
    PASS_TO_PASS (the existing test suite in the repository).

PREMISES (state what each patch does):
P1: Patch 1 modifies [file(s)] by [specific change description]
P2: Patch 2 modifies [file(s)] by [specific change description]
P3: The FAIL_TO_PASS tests check [specific behavior being tested]
P4: The PASS_TO_PASS tests check [specific behavior, if relevant]

ANALYSIS OF TEST BEHAVIOR:

For FAIL_TO_PASS test(s):
  Claim 1.1: With Patch 1 applied, test [name] will [PASS/FAIL]
             because [trace through the code behavior]
  Claim 1.2: With Patch 2 applied, test [name] will [PASS/FAIL]
             because [trace through the code behavior]
  Comparison: [SAME/DIFFERENT] outcome

For PASS_TO_PASS test(s) (if patches could affect them differently):
  Claim 2.1: With Patch 1 applied, test behavior is [description]
  Claim 2.2: With Patch 2 applied, test behavior is [description]
  Comparison: [SAME/DIFFERENT] outcome

EDGE CASES RELEVANT TO EXISTING TESTS:
(Only analyze edge cases that the ACTUAL tests exercise)

E1: [Edge case that existing tests exercise]
  - Patch 1 behavior: [specific output/behavior]
  - Patch 2 behavior: [specific output/behavior]
  - Test outcome same: [YES/NO]

COUNTEREXAMPLE (required if claiming NOT EQUIVALENT):
Test [name] will [PASS/FAIL] with Patch 1 because [reason]
Test [name] will [FAIL/PASS] with Patch 2 because [reason]
Therefore patches produce DIFFERENT test outcomes.

OR

NO COUNTEREXAMPLE EXISTS (required if claiming EQUIVALENT):
All existing tests produce identical outcomes because [reason]

FORMAL CONCLUSION:
By Definition D1:
- Test outcomes with Patch 1: [PASS/FAIL for each test]
- Test outcomes with Patch 2: [PASS/FAIL for each test]
- Since test outcomes are [IDENTICAL/DIFFERENT], patches are
  [EQUIVALENT/NOT EQUIVALENT] modulo the existing tests.

ANSWER: [YES/NO]
\end{verbatim}
\end{tcolorbox}

This template requires the agent to explicitly state premises about each patch, trace test execution for both patches, and provide either a counterexample (for non-equivalence) or a proof that no counterexample exists (for equivalence). The structured format ensures the agent cannot make unsupported claims about patch behavior.

\paragraph{Example: Semi-formal Analysis Discovers Name Shadowing (django-13670).}
This example demonstrates how semi-formal reasoning detects subtle bugs that surface-level reasoning misses.

\emph{Task:} Determine if two patches that fix 2-digit year formatting for years $<$ 1000 produce identical test outcomes.

\emph{Ground Truth:} NOT EQUIVALENT (Patch 1 calls wrong \texttt{format} function causing AttributeError)

\begin{tcolorbox}[colback=gray!5, colframe=gray!50, boxrule=0.3pt, breakable, title=Patch Comparison]
\small
\begin{verbatim}
# Patch 1 (FAILS):
def y(self):
    return format(self.data.year, "04d")[-2:]

# Patch 2 (WORKS):
def y(self):
    return '%02d' % (self.data.year % 100)
\end{verbatim}
\end{tcolorbox}

\textbf{Standard reasoning} (INCORRECT) concluded:
\begin{quote}
\emph{``Both patches fix the same bug. Patch 1 uses \texttt{format(self.data.year, '04d')[-2:]} which pads to 4 digits and takes the last 2, while Patch 2 uses modulo 100 and pads to 2 digits. For year 476: Patch 1 gives \texttt{'0476'[-2:] = '76'} and Patch 2 gives \texttt{'76'}. Both produce identical results.''}
\end{quote}

The agent assumed \texttt{format()} refers to Python's builtin, which would produce the expected behavior.

\textbf{Semi-formal reasoning} (CORRECT) mandated checking function definitions:

\begin{tcolorbox}[colback=gray!5, colframe=gray!50, boxrule=0.3pt, breakable]
\small
\begin{verbatim}
STEP 1: METHOD DEFINITIONS (read from source code)

For Patch 1, method y() calls format(). Checking module imports...

CRITICAL DISCOVERY: Module-level format function (line 340-343):
    def format(value, format_string):
        "Convenience function"
        df = DateFormat(value)
        return df.format(format_string)

STEP 2: FRAMEWORK SEMANTICS ANALYSIS

- Property 1: Module defines function named 'format' at line 340
- Property 2: Python name resolution: local -> module -> builtins
- Property 3: When y() calls format(), Python finds MODULE-LEVEL
  format() BEFORE the builtin format()
- Property 4: Module-level format(value, format_string) expects
  'value' to be a datetime/date object
- Property 5: format(476, '04d') creates DateFormat(476)
- Property 6: DateFormat(476).format('04d') accesses self.data.day
  but 476 (int) has no .day attribute

COUNTEREXAMPLE:
Test test_year_before_1000 will FAIL with Patch 1:
- format(476, '04d') -> DateFormat(476).format('04d')
- Raises AttributeError: 'int' object has no attribute 'day'

Test will PASS with Patch 2:
- '%02d' % (476 % 100) = '76' (no external function calls)

ANSWER: NO (patches are NOT equivalent)
\end{verbatim}
\end{tcolorbox}

The structured template required tracing the execution path through each patch. This led the agent to check the local \texttt{format} definition, read the complete function, and derive that passing an integer would cause AttributeError, uncovering that \texttt{format} is shadowed by a module-level function with completely different semantics than Python's builtin.

\section{Fault Localization Semi-formal Reasoning Template}
\label{app:fl-prompt}

The semi-formal template for fault localization enforces a structured reasoning process that traces from test semantics through code paths to divergence claims. The key insight is the PREMISE $\rightarrow$ CLAIM $\rightarrow$ PREDICTION chain: every prediction must trace back through a divergence claim to a specific test premise.

\paragraph{Semi-formal Prompt Template.}
The agent receives the failing test name, test method code, list of available source files, and list of passing tests. The structured reasoning template requires four phases:

\begin{tcolorbox}[
  title=Semi-formal Fault Localization Template,
  colback=gray!5,
  colframe=gray!75!black,
  fonttitle=\bfseries,
  breakable,
  enhanced
]
\begin{verbatim}
## Phase 1: Test Semantics Analysis
- What does the failing test method do step by step?
- What are the explicit assertions / expected exceptions?
- What is the expected behavior vs. the observed failure mode?
- State these as formal PREMISES:
  PREMISE T1: The test calls X.method(args) and expects [behavior]
  PREMISE T2: The test asserts [condition]
  ...

## Phase 2: Code Path Tracing
- Trace the execution path from the test's entry point into
  production code
- For each significant method call, document:
  METHOD: ClassName.methodName(params)
  LOCATION: file:line
  BEHAVIOR: what this method does
  RELEVANT: why it matters to the test
- Build a call sequence showing the flow from test -> production code

## Phase 3: Divergence Analysis
- For each code path traced, identify where the implementation
  could diverge from the test's expectations
- State divergences as formal claims:
  CLAIM D1: At [file:line], [code] would produce [behavior]
            which contradicts PREMISE T[N] because [reason]
  CLAIM D2: ...
- Each claim must reference a specific PREMISE and a specific
  code location

## Phase 4: Ranked Predictions
- Based on the divergence claims, produce ranked predictions
- Each prediction must cite the supporting CLAIM(s)
\end{verbatim}
\end{tcolorbox}

\paragraph{Structured Exploration Format.}
During agentic exploration, each file request and observation must follow a structured format:

\begin{tcolorbox}[
  colback=gray!5,
  colframe=gray!75!black,
  breakable,
  enhanced
]
\begin{verbatim}
### When requesting a file:

HYPOTHESIS H[N]: [What you expect to find and why it may
                  contain the bug]
EVIDENCE: [What from the test or previously read files
           supports this hypothesis]
CONFIDENCE: [high/medium/low]

### After reading a file:

OBSERVATIONS from [filename]:
  O[N]: [Key observation about the code, with line numbers]
  O[N]: [Another observation]

HYPOTHESIS UPDATE:
  H[M]: [CONFIRMED | REFUTED | REFINED] - [Explanation]

UNRESOLVED:
  - [What questions remain unanswered]
  - [What other files/functions might need examination]

NEXT ACTION RATIONALE: [Why reading another file, or why
                        enough evidence to predict]
\end{verbatim}
\end{tcolorbox}

This structured format forces the agent to: (1) form explicit hypotheses before exploring code, (2) document observations with precise line numbers, (3) update hypotheses based on evidence rather than intuition, and (4) justify each action in the exploration process. The explicit hypothesis tracking reduces the tendency to pattern-match on function names or make unsupported claims about code behavior.

\section{Case Study for Mockito\_8}
\label{app:mockito8}

Mockito\_8 involves a StackOverflowError caused by infinite recursion in Java generic type resolution.
The bug is in GenericMetadataSupport.java.
When processing a self-referential type like SelfRef<T extends SelfRef<T>>, the method registerTypeVariablesOn at line 80 overwrites a safe T → TypeVarBoundedType(T) mapping with a self-referential T → T, causing getActualTypeArgumentFor to recurse infinitely. The ground-truth fix is at line 79.

\textbf{Standard reasoning} (INCORRECT) predicted getActualTypeArgumentFor (lines 181-189), the crash site where the StackOverflowError manifests. The agent used only a single turn and all 3 predictions targeted the recursive method and its immediate callers. The reasoning correctly identified the infinite recursion but stopped at the symptom:

"StackOverflowError due to infinite recursion. When a TypeVariable maps to itself in contextualActualTypeParameters, getActualTypeArgumentFor recurses endlessly at line 185."

\textbf{Semi-formal reasoning} (CORRECT) correctly identified the root cause at Top-1. The 4-phase template drove a more thorough investigation across 2 turns, tracing back from the crash to its root cause.

\textbf{Phase 1: Test Semantics Analysis.} The agent established formal premises about the test:

\begin{tcolorbox}[colback=gray!5, colframe=gray!50, boxrule=0.3pt, breakable]
\small
\begin{verbatim}
PREMISE T1: The test typeVariable_of_self_type tests a method
where the return type is a type variable that references itself
(self-referential type, like <T extends Comparable<T>>).

PREMISE T2: Based on the GenericsNest interface,
two_type_params() returns T where <S, T extends S>.

PREMISE T3: The test would call
inferFrom(...).resolveGenericReturnType(...)
and make assertions about the result.
\end{verbatim}
\end{tcolorbox}

These premises guided the agent toward self-referential type patterns rather than immediately fixating on the crash site.

\textbf{Phase 2: Code Path Tracing.} The agent produced 10 formal observations (O1--O10) tracing the full execution path:

\begin{tcolorbox}[colback=gray!5, colframe=gray!50, boxrule=0.3pt, breakable]
\small
\begin{verbatim}
O1-O3: inferFrom(Type) -> FromClassGenericMetadataSupport ->
  resolveGenericReturnType(Method) creates TypeVariableReturnType

O4-O5: TypeVariableReturnType constructor calls
  readTypeParameters() -> registerTypeParametersOn

O6-O8: registerTypeVariableIfNotPresent stores
  boundsOf(typeVariable) -- for <S, T extends S>,
  T maps to TypeVarBoundedType(S) (safe, wrapped in BoundedType)

O9-O10: readTypeVariables() iterates bounds and calls
  registerTypeVariablesOn(type) for each, then
  registerTypeVariablesOn(getActualTypeArgumentFor(typeVariable))
\end{verbatim}
\end{tcolorbox}

The agent then traced multiple concrete type scenarios (\texttt{<S, T extends S>}, \texttt{K extends Comparable<K>}, \texttt{UpperBoundedTypeWithClass<E extends Number \& Comparable<E>>}) and verified each worked correctly, systematically ruling out false leads.

\textbf{Phase 3: Divergence Analysis.} The agent formulated and tested three hypotheses:

\begin{tcolorbox}[colback=gray!5, colframe=gray!50, boxrule=0.3pt, breakable]
\small
\begin{verbatim}
HYPOTHESIS H1: getActualTypeArgumentFor enters infinite
  recursion when a TypeVariable maps to itself.
  Evidence: no cycle detection at lines 181-189.

HYPOTHESIS H2: readTypeVariables at line 379 might receive
  a BoundedType and silently skip necessary registration.

HYPOTHESIS H3: extractRawTypeOf at line 405 could return null
  for an unregistered TypeVariable.
\end{verbatim}
\end{tcolorbox}

The agent then tested each hypothesis against the self-referential pattern \texttt{SelfRef<T extends SelfRef<T>{}>}:

\begin{tcolorbox}[colback=gray!5, colframe=gray!50, boxrule=0.3pt, breakable]
\small
\begin{verbatim}
Step 1: registerTypeParametersOn([T]) stores
  T -> TypeVarBoundedType(T) (safe).

Step 2: readTypeVariables() processes T.getBounds() =
  [SelfRef<T>], calls registerTypeVariablesOn(SelfRef<T>).
  This is a ParameterizedType -- rawType = SelfRef,
  typeParameters = [T_SelfRef],
  actualTypeArguments = [T_current].
  But T_SelfRef and T_current are the SAME TypeVariable!
  So contextualActualTypeParameters.put(T, T)
  -- mapping T to itself!
  This OVERWRITES the existing T -> TypeVarBoundedType(T)!

Step 3: getActualTypeArgumentFor(T) looks up T -> gets T
  (a TypeVariable!) -> recurses -> INFINITE RECURSION!
\end{verbatim}
\end{tcolorbox}

The agent verified this does \emph{not} happen for non-self-referential cases like \texttt{UpperBoundedTypeWithClass<E extends Number \& Comparable<E>{}>}: there, \texttt{registerTypeVariablesOn(Comparable<E>)} maps \texttt{T\_Comparable} $\to$ \texttt{E}, but \texttt{T\_Comparable} $\neq$ \texttt{E} (different declarations), so \texttt{E}'s existing \texttt{BoundedType} mapping is preserved.

\textbf{Phase 4: Ranked Predictions.} The agent produced two formal claims and 5 ranked predictions:

\begin{tcolorbox}[colback=gray!5, colframe=gray!50, boxrule=0.3pt, breakable]
\small
\begin{verbatim}
CLAIM D1: At lines 73-81, registerTypeVariablesOn overwrites
  T -> TypeVarBoundedType(T) with T -> T for self-referential
  types, causing infinite recursion.

CLAIM D2: At lines 181-189, getActualTypeArgumentFor has no
  cycle detection.

Rank 1 (high): registerTypeVariablesOn lines 73-81
  -- the root cause (overwrite at line 80). [CORRECT]
Rank 2 (high): getActualTypeArgumentFor lines 181-189
  -- the crash site (alternative fix location).
Rank 3 (medium): readTypeVariables lines 375-379
  -- the caller that triggers the overwrite.
Rank 4 (low): WildCardBoundedType.equals lines 600-606
  -- an unrelated bug also found during analysis.
Rank 5 (low): extractRawTypeOf lines 400-406
  -- potential NPE from incomplete registration.
\end{verbatim}
\end{tcolorbox}

\paragraph{Why the structured template mattered.}
Without Phase 2's systematic code path tracing, the agent in standard mode stopped at the crash site, since \texttt{getActualTypeArgumentFor} is where the StackOverflowError manifests and is the obvious answer.
The semi-formal template forced the agent to trace the full registration pipeline (O1--O10), test multiple type scenarios, formulate competing hypotheses (H1--H3), and ultimately discover that the overwrite at line 80 is the root cause, not the recursive method that crashes.
The agent explicitly tested \texttt{SelfRef<T extends SelfRef<T>{}>} against \texttt{UpperBoundedTypeWithClass<E extends Number \& Comparable<E>{}>} to confirm why self-referential types uniquely trigger the bug, a differential analysis that standard mode never attempted.

\section{Code Question Answering Details}
\label{app:rubber-duck}

\paragraph{Semi-formal Prompt Template.}
The structured template used for semi-formal reasoning in code question answering requires the agent to fill in the following sections:

\begin{tcolorbox}[
  title=Semi-formal Reasoning Template,
  colback=gray!5,
  colframe=gray!75!black,
  fonttitle=\bfseries,
  breakable,
  enhanced
]
\begin{verbatim}
FUNCTION TRACE TABLE:
| Function/Method | File:Line | Parameter Types | Return Type | Behavior (VERIFIED) |
|-----------------|-----------|-----------------|-------------|---------------------|
| [function1]     | [file:N]  | [param types]   | [ret type]  | [ACTUAL behavior]   |

DATA FLOW ANALYSIS:
Variable: [key variable name]
- Created at: [file:line]
- Modified at: [file:line(s), or 'NEVER MODIFIED']
- Used at: [file:line(s)]

SEMANTIC PROPERTIES:
Property 1: [e.g., 'HashMap is mutable']
- Evidence: [specific file:line]

ALTERNATIVE HYPOTHESIS CHECK:
If the opposite answer were true, what evidence would exist?
- Searched for: [what you looked for]
- Found: [what you found - cite file:line]
- Conclusion: [REFUTED / SUPPORTED]

<answer>[Final answer with explicit evidence]</answer>
\end{verbatim}
\end{tcolorbox}

This template forces the agent to document evidence systematically before committing to an answer, reducing the tendency to guess based on function names.

\paragraph{Example: Semi-formal vs.\ Standard Reasoning.}
Consider a question asking about the difference between two similar API calls. With standard reasoning, the agent might guess based on naming conventions. With semi-formal, the agent must:
\begin{enumerate}
    \item Locate and read both function implementations
    \item Document their exact signatures and behavior in the trace table
    \item Identify semantic differences with specific file:line evidence
    \item Check whether the opposite conclusion is supported by evidence
\end{enumerate}

We present two concrete examples: one where semi-formal reasoning succeeds by forcing explicit verification, and one where it fails despite thorough analysis.

\paragraph{Success Case: Verifying Code Invariants (cpp\_3).}
\emph{Question:} ``Is there a difference between using \texttt{m\_sliceTypeStrMap.at(m\_sliceType)} vs \texttt{m\_sliceTypeStrMap[m\_sliceType]}?''

\textbf{Standard reasoning} (79\% score) correctly identified that \texttt{at()} is required for const-correctness, but stated: ``it provides proper error handling if an invalid SliceType is somehow passed''---implying non-existent keys could occur, which earned a deduction for making an irrelevant consideration seem important.

\textbf{Semi-formal reasoning} (100\% score) produced:

\begin{tcolorbox}[colback=gray!5, colframe=gray!50, boxrule=0.3pt, breakable]
\small
\begin{verbatim}
FUNCTION TRACE TABLE:
| Function             | File:Line               | Behavior (VERIFIED)                 |
|----------------------|-------------------------|-------------------------------------|
| toggleSlicingOptions | QtReflEventView.cpp:99  | Sets m_sliceType to SliceType enum  |

DATA FLOW ANALYSIS:
Variable: m_sliceTypeStrMap
- Created at: QtReflEventView.h:69-73 (initialized with 4 SliceType mappings)
- Modified at: NEVER MODIFIED

Variable: m_sliceType
- Modified at: QtReflEventView.cpp:99 (only assigned from SliceType enum)

SEMANTIC PROPERTIES:
Property: The map is pre-populated with all valid SliceType values
- Evidence: Header shows UniformEven, Uniform, Custom, LogValue in map
- Evidence: toggleSlicingOptions only assigns these 4 values
\end{verbatim}
\end{tcolorbox}

The template forced the agent to trace both the map initialization AND the \texttt{m\_sliceType} assignments, revealing they use exactly the same enum values---proving non-existent keys are impossible.

\paragraph{Failure Case: Confident Wrong Answer (py\_5).}
\emph{Question:} ``Do we need the check for `if not build\_directory'?''

The correct answer is that the check is \emph{not} needed because the \texttt{Runner} constructor already handles this case. However, semi-formal reasoning (57\% score) produced:

\begin{tcolorbox}[colback=gray!5, colframe=gray!50, boxrule=0.3pt, breakable]
\small
\begin{verbatim}
FUNCTION TRACE TABLE:
| Function            | File:Line           | Behavior (VERIFIED)                     |
|---------------------|---------------------|-----------------------------------------|
| setup_build         | build_manager:1397  | Returns Build object on success         |
| setup_system_binary | build_manager:1387  | Creates SystemBuild with SYSTEM_BINARY_DIR |
| SystemBuild.setup   | build_manager:1001  | Always returns True                     |

SEMANTIC PROPERTIES:
Property: setup_build can return truthy value while BUILD_DIR is empty
- Evidence: SystemBuild uses get_value('SYSTEM_BINARY_DIR', '') defaulting to ''

<answer>
Yes, the check IS NEEDED because setup_build() can return truthy
even when BUILD_DIR is empty string.
</answer>

CONFIDENCE: HIGH
\end{verbatim}
\end{tcolorbox}

\begin{table}[h]
\centering
\caption{Code QA accuracy by language (Opus-4.5, Semi-formal)}
\label{tab:rubber-duck-lang}
\begin{tabular}{lcc}
\toprule
\textbf{Language} & \textbf{Questions} & \textbf{Accuracy} \\
\midrule
C++ & 5 & 88.0\% \\
Python & 5 & 87.5\% \\
Java & 5 & 85.5\% \\
\bottomrule
\end{tabular}
\end{table}

The agent thoroughly traced five functions and found a real edge case where \texttt{BUILD\_DIR} could be empty. However, it \emph{missed} that \texttt{Runner}'s constructor also checks \texttt{build\_directory} (indirectly through \texttt{find\_fuzzer\_path}, which returns \texttt{None} if \texttt{build\_directory} is falsy, triggering a \texttt{CorpusPruningException}). This illustrates how deeper reasoning can lead to \emph{more confident wrong answers} when the agent follows a plausible-but-incorrect chain without verifying all downstream code paths.

\paragraph{Per-Language Breakdown.}
Table~\ref{tab:rubber-duck-lang} shows accuracy varies by programming language, with Java questions being most challenging due to their reliance on framework-specific semantics.

\end{document}